# The Impact of Changes to Daylight Illumination level on Architectural Experiences in Offices Based on VR and EEG

Pegah Payedar-Ardakani[1], Yousef Gorji-Mahlabani[2*], Abdul Hamid Ghanbaran[3], Reza Ebrahimpour[4]

[1] PhD candidate, Faculty of Architecture and Urbanism, Imam Khomeini International University, Qazvin, Iran.
[2*] Full Professor, Faculty of Architecture and Urbanism, Imam Khomeini International University, Qazvin, Iran.
[3] Associate Professor, Faculty of Architecture and Urban Planning, Shahid Rajaee Teacher Training University, Tehran, Iran.
[4] Full Professor, Centre for Cognitive science, Institute for Convergence Science and Technology, Sharif University of Technology, Tehran, Iran.

[*] Corresponding Author: Yousef Gorji-Mahlabani, *Email: gorji@arc.ikiu.ac.ir*

**Abstract**

This study investigates the influence of varying daylight illuminance levels on architectural experiences in a virtual office environment. Integrating subjective assessments and electroencephalogram (EEG) data, we aim to comprehensively understand how illuminance impacts emotional and neurophysiological responses. The experiment exposes participants to nine illuminance levels, ranging from 66 to 1500 lux, to discern optimal conditions for different architectural experiences. Subjective evaluations, gathered via questionnaires, required participants to rate the perceived pleasantness, interest, excitement, calmness, and spaciousness. Simultaneously, EEG data was recorded to analyze neurophysiological changes associated with different illuminance conditions. The results showed that the power of the $\alpha$ and $\theta$ band at channels in the parietal and centro-parietal regions had statistically significant differences under nine daylight illuminance levels. Illuminance changes activated brain regions associated with cognitive functions. The relative power of $\alpha$ band at the parietal region were negatively correlated with the subjective evaluation, and was relatively low at levels between 300 and 900 lux, indicative of heightened pleasantness, interest, spaciousness, with the peak experience occurring in the 700 to 900 lux range. Illuminance exceeding 1300 lux, conversely, correlated with increased absolute power of the alpha band, suggesting heightened arousal, while levels below 300 lux led to decreased arousal and increased calmness. By integrating neurophysiological measures with subjective evaluations, we gain a nuanced understanding of how illuminance affects emotional and cognitive responses. This research may provide a reference for the selection of office illuminance levels for employees during high-intensity mental work and rest.

# 1. Introduction

Given that individuals dedicate a significant amount of time indoors (Höppe, 2002), it becomes crucial to recognize and measure the connection between the interior-built environment and the human experience. The human experience within the built environment pertains to the mental state of individuals influenced by their surroundings, manifesting in psychological, physiological, and emotional dimensions. This experience notably impacts the cognitive and physical well-being of individuals, influencing moods, comfort levels, and interactions with the surroundings (Eberhard, 2009).

The term "user experience" design originated in the field of industrial design, introduced by Don Norman in 1995. It denotes the enhancement of the interaction between the product and the user to optimize the product's performance. In 2015, the term gained increased recognition with the release of Kim's book, *Design for Experience: Where Technology Meets Design*, and Peter Benz's book, *Experience Design: Concepts and Case Studies* (Chowdhury et al., 2020; Noguchi et al., 2022). "Architectural experiences" in the built environment are influenced by three major neural systems: knowledge-meaning, emotion-valuation, and sensorimotor systems (Chatterjee & Vartanian, 2014; Coburn et al., 2017; Jam et al., 2022). According to this, Coburn et al. (2022) proposed that architectural encounters generate three broad categories of psychological experiences: cognitive judgments associated with knowledge-meaning systems, emotional responses derived from emotion-valuation systems, and behavioral-motivational responses linked to sensorimotor activation (Coburn et al., 2020). Additionally, presented a conceptual model outlining component that influence the aesthetic experience of architecture, organized into six sections: perceptual, motivational-sensory, cognitive, emotional, and behavioral (Moosavian, 2022).

Daylighting, as a crucial architectural feature, is easy to notice by people and holds significant power to shape the human experience (Ergan et al., 2018; Gorji Mahlabani et al., 2011, 2019). The quality of daylighting impacts occupants through both biopsychological and psychological processes, influencing human mood, cognitive performance, social behavior, overall well-being, task performance, daytime alertness, and the duration of sleep (Boubekri et al., 2020; de Vries et al., 2018; Ghanbaran et al., 2018; Küller et al., 2006; Spengler, 2012; Veitch et al., 2013).

In past studies, researchers have predominantly evaluated the architectural experience of daylight through surveys, experiments, or a blend of both methods (Payedar-Ardakani et al., 2023). Alongside interviews, employed in certain surveys like (Venugopal et al., 2020), other researchers have employed two types of questionnaires to assess emotional experiences: The Positive and Negative Affect Schedule (PANAS) and the PAD model. The utilization of the PAD model was initially validated in daylight studies by (Boubekri et al., 1991). Their findings revealed that the size of sunlit areas had a substantial impact on the feelings of relaxation and excitement among predominantly female office workers.

Moreover, the Experimental method stands out as one of the most crucial approaches for measuring physiological and neurophysiological experiences, often employed in conjunction with neuroscience and biometric sensors. Ergan et al. (2019) introduced an integrated method utilizing biometric sensors such as electroencephalogram (EEG), Galvanic Skin Response (GSR), and photoplethysmogram (PPG) to quantify the sense of stress and anxiety (assessed through a PANAS-related questionnaire) in structured user experiments conducted within a visualization laboratory using alternate Virtual Environments (VEs). The virtual environments were configured by manipulating a set of architectural design features, including daylighting (Zou & Ergan, 2019).

Chamilothori et al. (2019) delved into the impact of façade and sunlight pattern geometry on occupants' physiological responses in Virtual Reality (VR) and their subjective perception. The assessment was based on the PAD model, encompassing dimensions of pleasantness, calmness, and excitement for emotional evaluation, along with factors such as interest, complexity, spaciousness, satisfaction with exterior view, amount of brightness. In this study, Electrodermal Activity (EDA), blood volume pulse data (BVP), and heart rate (HR) were gathered using an Empatica E4 wristband device. The investigation involved the study of irregular and regular shadings, blinds with clear skies, and the penetration of direct light in two distinct scenarios (Chamilothori et al., 2019).

The researchers extended their investigation to other variables related to regional differences in the perception of daylit scenes (Chamilothori et al., 2022a). Additionally, they explored the effects of Sky Type, Space Function, and latitude in two separate manuscripts (Chamilothori et al., 2022b). In a study by Moscoso et al. (2021), the same eight variables were employed to assess the effects of window size on subjective impressions of daylit spaces at high latitudes, utilizing virtual reality (Moscoso et al., 2021). Similar scales have been utilized in some other studies as well. In their work, Kong et al. (2022) explored the impacts of aperture design, window size, and sky condition on both subjective and physiological responses within immersive virtual reality scenes. Participants rated degrees of eight adjectives, following the approach of Chamilothori et al.'s studies, while physiological data, including heart rates and EEG, were collected. The study revealed that both aperture designs and sky types exerted influences on subjective responses. Specifically, the presence of a large window enhanced beta oscillations and beta power in the right prefrontal lobe area, and clear sky conditions attenuated theta rhythm in the prefrontal lobe areas (Kong et al., 2022). Furthermore, Pastore and Andersen (2022) utilized both monitoring and the rating of four adjectives, drawing on the methodology from Chamilothori et al.'s studies. Their objective was to assess the impact of façade and space design on the indoor experience of building occupants. The study examined a ventilated façade, a double-skin façade with colored silk-printed and chrome-based patterns, and floor-to-ceiling fixed windows (Pastore & Andersen, 2022).

Eye and head tracker devices are other sensors for investigating the architectural experience in daylight spaces. Gökaslan & Erkan (2020) employed online eye tracking techniques to examine the impact of different window sizes on the cognitive experience of individuals. The results indicated that the window size within a space can influence people's focus, consequently altering user interactions with that space (Gökaslan & Erkan, 2020). Fathy et al. (2023) utilized head tracking measurements to identify areas of interest within a space under varying daylight conditions, incorporating virtual reality and machine learning into their methodology (Fathy et al., 2023).

Despite the extensive exploration of factors associated with daylight in various architectural experiences in these studies, it's noteworthy that most of them do not incorporate any lighting measurements. For instance, despite the significance of overall illuminance level as one of the crucial qualities of light to consider (Mostafavi et al., 2023), there exists a deficiency in objective methods for characterizing subjective experiences and emotions in daylight illumination studies. In certain studies, such as (Jakubiec et al., 2021; Mirdamadi et al., 2023), only subjective assessments were employed. While subjective evaluation is a valuable approach for studying human emotions and experiences, a commonly identified challenge is the substantial disparities in surveyed data, indicating biases or individual differences among observers. In addressing this issue, we established correlations between objective and subjective assessments using calibrated point-in-time daylighting models and images of an office, where occupants are virtually seated and exposed to different illuminance levels.

**EEG and Illumination evaluation:**

EEGs have been utilized to measure electrical activity in the brain (Ebrahimpour et al., 2023; Sarailoo et al., 2022) and explore the impact of design feature configurations, such as light conditions, on human psychology and cognitive load assessment (Li et al., 2020) . Previous research has indicated that EEG sensors demonstrate greater effectiveness compared to other sensors when it comes to classifying human experience in alternately designed spaces (Zou & Ergan, 2021). In the study conducted by Castilla et al. (2023), this approach was employed to investigate whether different illumination levels might impact university students' memories, as another cognitive function. Forty subjects engaged in a psychological memory task within three virtual classrooms with illuminance levels of 100, 300, and 500 lux. The findings revealed that, based on neurophysiological responses, participants' performance deteriorated, and their neurophysiological activation decreased with increasing illuminance (Castilla et al., 2023).

Oh et al. (2023) employed EEG to examine the impact of illumination levels at 150 and 1500 lux on the relative power of the beta and theta bands. The results indicated that there was no significant correlation between illumination levels and the power change of frequency bands (Oh et al., 2023). Fu et al. found evidence suggesting that the mean power spectral density of the alpha band at 300 lux is the highest, indicating lower attention and concentration, making it suitable for rest spaces (Fu et al., 2023). Armanto et al.'s experiment results showed that illumination at 1000 lux increased the absolute power of the alpha band compared to 300 lux (Armanto et al., 2009). Min et al. observed that with the increase in light intensity from 150 to 700 lux, the power of the alpha band in the parietal lobe of the brain decreases (Min et al., 2013). Additionally, Park et al. discovered a decrease in theta band activity in the frontal lobe of the brain with increasing illumination level (Park et al., 2013).

Indeed, it has been demonstrated that neurophysiological recording tools like EEG can be valuable for quantifying cognitive-emotional processes. However, there is a limited number of articles investigating the relationship between measured daylight illuminance levels and their effects on architectural experiences using EEG.

As suggested by Shan et al., EEG can not only elucidate and correlate the results obtained through questionnaire-based and task-based methods but also offer a more objective and potentially sensitive means to assess the impact of daylight on occupants (Shan et al., 2019). Consequently, the objective of this study is to analyze the impact of different illuminance levels of daylighting on the architectural experience of employees. This relationship is comprehensively examined through employees' perceptual and emotional responses (utilizing a Subjective Rating Questionnaire) as well as their neurophysiological responses measured through EEG, using virtual reality.

## 2. Material and methods

### 2-1. Preparation of VEs and Daylight conditions

For the study, an office room located on the fifth floor of a building was chosen as the test room. The room featured internal dimensions of 4.20 m × 3.30 m with a floor-to-ceiling height of 2.90 m and a south-facing window. To measure the illumination conditions within the test room and capture 360-degree panorama photos, a calibrated LX-102 Lutron Luxmeter and an INSTA360 One camera were utilized. A 1 × 1 square matrix of illuminance measuring points was arranged on the floor with a distance of 1 m between each measuring point. The camera and luxmeter were positioned horizontally over a table with a height of 0.85 m. The camera's height on the table was set at the observer's eye level when sitting (120 cm) and placed as close as possible to the luxmeter. Horizontal illumination measurements and photo capture were repeated at each measuring point.

High Dynamic Range Images (HDRI) were generated by combining five Low Dynamic Range Images (LDRI) with different exposure values. The lowest sensitivity (ISO) of 100 was employed to minimize noise in the HDRI, with a fixed white balance of 5000 K. The Reinhard tone mapping, using Luminance-hdr software version 2.6.0 (gamma of 2.2 and key value of 0.01) (Abd-Alhamid et al., 2019), was applied to the HDRI images. This tone mapping aimed to create contrast values similar to the real environment and compatible with the VR head-mounted display. The images were captured in July between 9:00 a.m. and 19:00 p.m. on two days under clear sky conditions. No additional light sources or reflective surfaces were present in the room to avoid glare. Nine specific conditions were selected for testing with illumination levels of 66, 173, 306, 438, 620, 790, 1011, 1220, and 1395 lux. Table 1 provides a visual representation of the selected 360 stereoscopic images in each daylighting mode, respectively.

**Table 1.** The selected 360˚ stereoscopic images in each illumination level.

| Level 1: 66 Lux (< 100) | Level 2: 173 Lux (100-300) | Level 3: 306 Lux (300-500) |
|---|---|---|
| 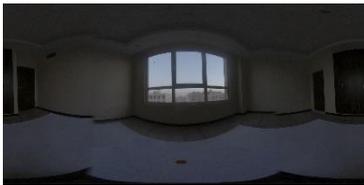 | 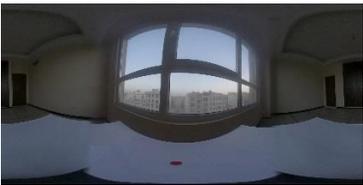 | 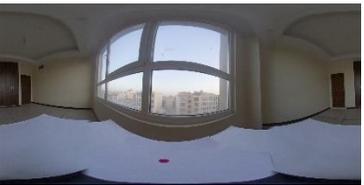 |
| Level 4: 438 Lux (300-500) | Level 5: 620 Lux (500-700) | Level 6: 790 Lux (700-900) |
| 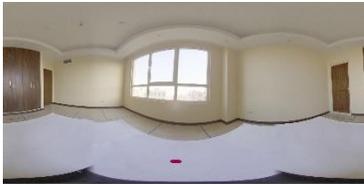 | 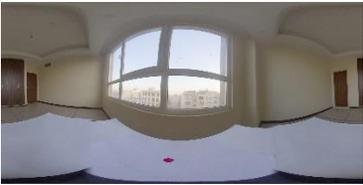 | 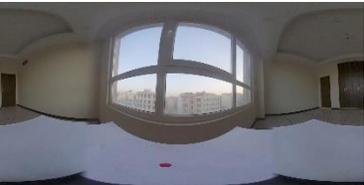 |
| Level 7: 1011 Lux (900-1100) | Level 8: 1220 Lux (1100-1300) | Level 9: 1395 Lux (1300-1500) |

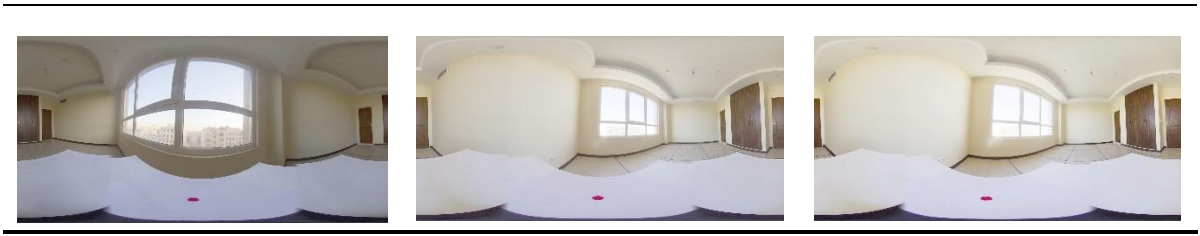

Finally, the generated images were imported into the *Unity* gaming engine (version 2019.4.40f1) using the *C-Sharp* (C#) programming language. These images were applied as textures to an identical concentric sphere within Unity, specifying a gamma color space and an Unlit two-sided material, ensuring it remained unaffected by light sources. To control the view direction in the scene, a virtual camera from the *OpenVR* system was placed at the center of this sphere. The direction of the view in the *Unity* scene was then manipulated through head movements in the VR headset. This spherical representation was subsequently projected to the corresponding eye in *Unity*, enabling the creation of a fully stereoscopic scene from the participant's viewpoint.

To validate the similarity between the illuminance from the VR display and the horizontal illuminance measured in the real environment from the same viewing position, the illuminance received at the eye (10 mm from the lens) was measured using a Luxmeter in a completely dark environment (devoid of any other source of illumination). The measured illuminance was 620 lux compared to 657 lux when the scene is displayed (~1.06 times), which is below the threshold of 1.50 for a noticeable variation in illuminance, as defined by the Comite Europeen de Normalisation (CEN) in 2011 (ISO, 2011). The immersive 360º images were presented using an HTC Vive Pro head-mounted display. This VR device features dual AMOLED 3.5" diagonal displays, boasting a combined resolution of 2880 × 1600 pixels (1440 × 1600 pixels per eye). With a refresh rate of 90 Hz and a nominal field of view of 110º, it enhances the immersive viewing experience.

### 2-3. Measures

This study gathered self-reported assessments through a subjective questionnaire and collected neurophysiological data, including electroencephalogram (EEG).

#### 3-3-1. Subjective Questionnaire

Each participant was required to complete two questionnaires. The first questionnaire included demographic information such as gender, age, eyesight, education, grade, and the daily duration spent in offices or for study. The second questionnaire, following Moscoso and Chamilothori's study (Chamilothori et al., 2022a, 2022b), consisted of six questions. Participants were asked to provide responses on a scale of 0 (not at all) to 11 (very) after experiencing each illuminance level condition. This virtual questionnaire, presented as visual and auditory, required participants to rate the perceived pleasantness, interest, excitement, calmness, and spaciousness of the space. The scores obtained for each illumination level in these variables were utilized in our data analysis. Prior to the experiment, the meanings of these variables were explained to the subjects. It is important to note that the two questions in the referenced questionnaire, related to brightness and satisfaction with the amount of view in the space, were excluded as they fell outside the scope of our study's objectives. Table 2 provides an overview of the subjective assessments collected in this study.

Table 2. overview of SAM questionnaire collected in the study.

| SAM Questionnaire |
|---|
| How pleasant is this space? |
| How interesting is this space? |
| How exciting is this space? |
| How calming is this space? |
| How complex is this space? |
| How spacious is this space? |

3-3-2. Neurophysiological Evaluation

In this study, the EEG signals of the participants were collected as neurophysiological evaluation. The g.HIamp electroencephalogram device from g.tec company was employed for brain mapping. EEG data were recorded from 32 active electrodes, placed according to the international 10/20 system, at a sampling rate of 512 Hz. To ensure optimal contact with the scalp, participants were instructed to wash and dry their hair before the test, aiming to reduce the impact of oil and skin keratin on conductivity.

**2-4. Experimental Procedure**

All experiment sessions were conducted at the National Brain Mapping Laboratory (NBML), located in the same physical location, from the end of May to September in 2023. The indoor air temperature during the experiment ranged between 22 ºC to 26 ºC. After the preparation phase, a pilot experiment involving 3 participants was conducted to assess potential discomfort during simultaneous VR-EEG measurement, identify any artifacts or noises, and ensure seamless data acquisition. Fig. 1 illustrates the two-step experiment procedure, comprising EEG measurement and the subsequent survey.

After entering the lab, participants were briefed on the experimental purposes and procedures, including their right to withdraw from the experiment at any time. Once participants agreed to participate, they completed the startup questionnaire. Following an adequate break, researchers, with the assistance of a skilled laboratory operator, fitted participants with EEG sensors. The HTC Vive Pro Eye HMD was then placed over the participant's head comfortably, ensuring a clear view of the virtual environment. Impedance of all electrodes was double-checked, and the status of EEG data recording from participants was examined before and after wearing the VR equipment. Additionally, researchers ensured the absence of incoming noises and the stability of EEG before starting the experiment.

Initially, participants viewed a gray screen in the VR, maintaining a comfortable state without any stimuli, while 2 minutes of baseline EEG data with eyes open were collected. The experimental trial commenced with an introductory gray-scale situation (depicting the office room in gray-scale). Participants were instructed to envision working in this office, minimize body movement, and remain silent throughout the experiment, focusing on the red point on their virtual table. The introductory situation was displayed for 40 seconds. Subsequently, participants were sequentially exposed to nine different illumination conditions, ranging from dim to bright. Before presenting each scene, a gray color scene was displayed in VR for 5 seconds to ensure chromatic adaptation.

For each condition, the participant first experienced it for 10 s and then the researchers collected their verbal responses to the seven questions of the evaluation questionnaire, which took around 70-80 s. The words "pleasant, interesting, exciting, calming, complex & spacious" in the questions were displayed in red in the VR and played audibly. Finally, after completing all nine conditions, the researcher assisted the participant with removing the EEG and VR equipment. The entire experiment lasted between 60 and 70 min and the total time participants spent in the VR was around 16–25 min (Fig. 2). The researcher observed the psychological data throughout each participant's entire experiment.

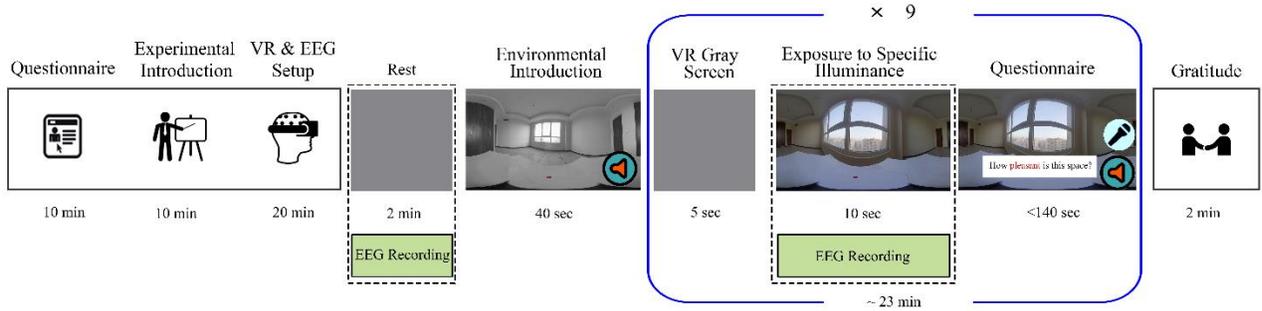

Fig.1. Experimental procedure.

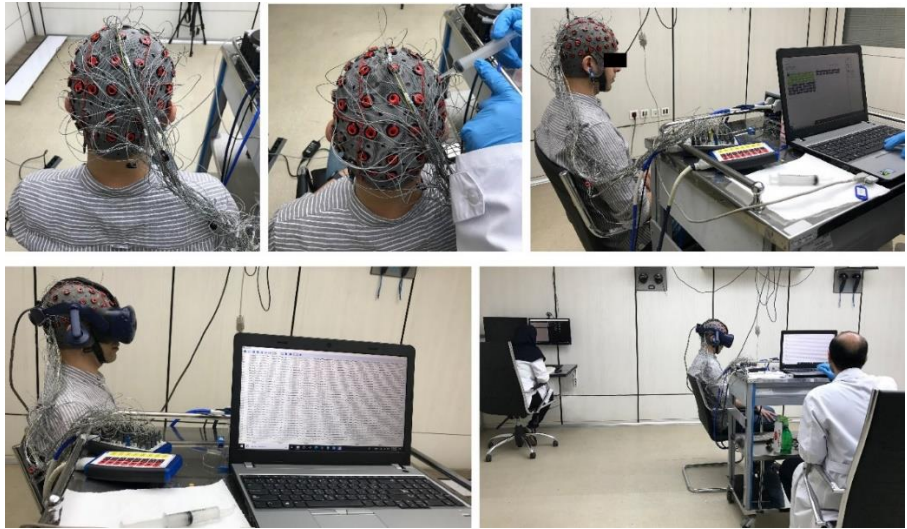

Fig. 2. The setting up process and a study participant wearing the VR head-mounted display and EEG cap during the task process.

### 2-5. Statistical Analysis

The content and structure of the data for analysis in the experiment are outlined in Fig. 3. The normality of the data was assessed using the Shapiro–Wilk test, and the homogeneity of variances was examined with the Levene test. One-way ANOVA was employed to analyze the variability of the data across different illuminance levels. The Pearson correlation coefficient method was utilized to examine the closeness of the relationship between variables. All data processing was carried out using SPSS 27.0 (IBM Corp., 2020).

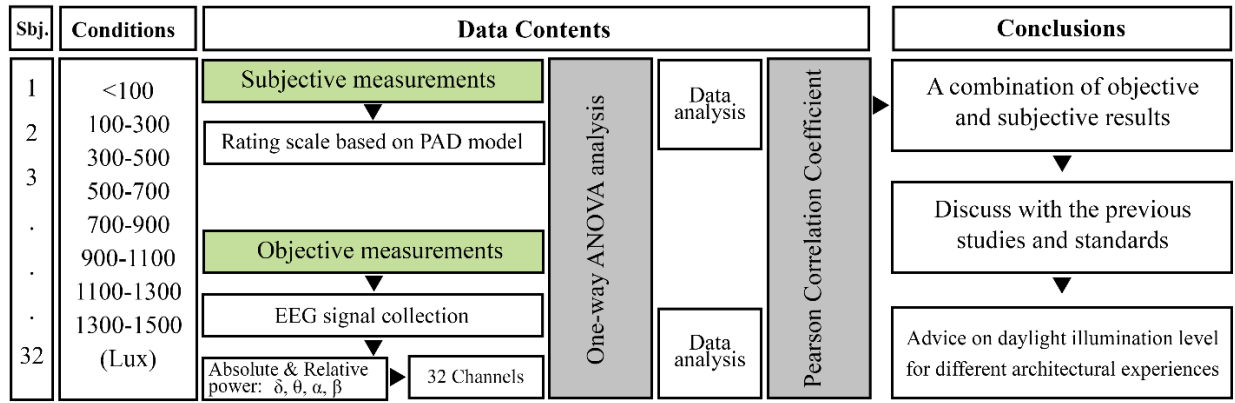

Fig.3. The content and structure of the data analysis.

## 2-6. EEG Preprocessing

The raw EEG signals were pre-processed using MATLAB version 2016b and EEGLAB 2021. The signals underwent filtering with a frequency range of 1 to 40 Hz using the FIR filter for analysis. Noisy time intervals were eliminated through visual rejection, and noisy channels were removed and interpolated. Independent Component Analysis (ICA) and Automated Artifact Rejection (ASR) were applied to further remove artifactual components, including blinking, body, and muscle movements from the EEG data. The time division of trials was conducted using existing triggers with a duration of 10 seconds (representing the display duration of each scene). The time domain signals were then transformed into the frequency domain using Fast Fourier Transform (MATLAB Pweltch function). The resulting frequency bands were categorized into four bands: $\delta$ (1 Hz~4 Hz), $\theta$ (4 Hz~8 Hz), $\alpha$ (8 Hz~12 Hz), and $\beta$ (12 Hz~30 Hz). Absolute and relative power values of the EEG signal for each participant's channels during each scene were extracted. Relative power indicates the contribution of each sub-frequency power relative to the total power, potentially reducing differences between subjects compared to absolute power (Poza et al., 2013; Yu et al., 2022).

## 2-7. Participants

To address the experiment's aim, a total of 32 physically and mentally healthy adults (16 male, 16 female) were selected as experimental subjects using a simple random sampling method. The subjects ranged from 20 to 40 years of age, with a mean age of 27.25 years and a standard deviation of 6.294. Basic information about the subjects is presented in Table 3. All participants had normal or corrected-to-normal vision and did not have any psychiatric or neurological disorders, head trauma, history of smoking, alcohol addiction, or drug abuse. The subjects ensured they had at least 8 hours of sleep on the night before the experiment. The experimental protocols received approval from the Research Ethics Committees of Iran University of Medical Sciences (IR.IUMS.REC.1401.1033). Prior to participating in the experiment, each participant provided written informed consent, and their consent was obtained and approved in advance.

**Table 3.** Basic information about subjects.

|  |  | Male | Female | Total |
|---|---|---|---|---|
| Frequency | N | 16 | 16 | 32 |
| Age | Max | 40 | 39 | 40 |

|  | Min | 20 | 20 | 20 |
|---|---|---|---|---|
|  | Mean | 27.13 | 27.38 | 27.25 |
|  | SD | 7.013 | 5.714 | 6.294 |

## 3. Results
### 3-1. The Impact of Different Illuminations on Subjective measurements

One-way Analysis of Variance (ANOVA) was conducted to examine the impact of illuminance levels on pleasantness, calmness, excitement and arousal, interest, perception of complexity, and spaciousness. As indicated in Table 4, there is a statistically significant difference at the P<0.05 level in the scores of pleasantness (F=9.161, P<0.0001), excitement (F=5.832, P<0.0001), interest (F=4.748, P<0.0001), and spaciousness (F=7.400, P<0.0001) across the nine illumination groups. Consequently, the relationship between these variables has been further analyzed in conjunction with the results of brain mapping.

Table 4. Participants' SAM ratings for nine-level illuminated VR images.

| Dependent Variable | | Sum of Squares | df | Mean Square | F | Sig. |
|---|---|---|---|---|---|---|
| pleasantness | Between Groups | 3.371 | 8 | .421 | 9.161 | .000 |
|  | Within Groups | 12.007 | 261 | .046 |  |  |
|  | Total | 15.378 | 269 |  |  |  |
| calmness | Between Groups | .484 | 8 | .060 | 1.196 | .302 |
|  | Within Groups | 13.204 | 261 | .051 |  |  |
|  | Total | 13.688 | 269 |  |  |  |
| interest | Between Groups | 1.908 | 8 | .238 | 4.748 | .000 |
|  | Within Groups | 13.108 | 261 | .050 |  |  |
|  | Total | 15.016 | 269 |  |  |  |
| excitement | Between Groups | 2.481 | 8 | .310 | 5.832 | .000 |
|  | Within Groups | 13.882 | 261 | .053 |  |  |
|  | Total | 16.364 | 269 |  |  |  |
| complexity | Between Groups | .643 | 8 | .080 | 1.459 | .173 |
|  | Within Groups | 14.371 | 261 | .055 |  |  |
|  | Total | 15.013 | 269 |  |  |  |
| spaciousness | Between Groups | 3.763 | 8 | .470 | 7.400 | .000 |
|  | Within Groups | 16.591 | 261 | .064 |  |  |
|  | Total | 20.355 | 269 |  |  |  |

### 3-2. The Impact of Different Illuminations on EEG

One-way ANOVA was utilized to assess whether illuminance affects subjects' neurophysiological responses. The absolute and relative power of Delta, Theta, Alpha, and Beta bands at each of the 32 channels in different illuminance environments were analyzed, and the significant results of the one-way ANOVA were visually represented on electrode layout maps in Figures 4 and 5, with red color indicating stronger significance. Changing the illuminance level did not have a significant effect on the power of delta and beta bands. For absolute power values in the Alpha band, significant differences were found at the CP1 electrode (p = 0.030, F=2.18), CP2 electrode (p = 0.026, F=2.23), P3 electrode (p = 0.048, F=1.998), Pz electrode (p = 0.021, F=2.310), P8 electrode (p = 0.023, F=2.286), and O1 electrode (p = 0.041, F=2.061) in the parietal region.

For relative power values of the alpha band, significant differences were found at 15 channels containing FC6 electrode (p = 0.007, F=2.707), C3 electrode (p = 0.022, F=2.290), Cz electrode (p = 0.014, F=2.461), C4 electrode (p = 0.010, F=2.566), CP5 electrode (p = 0.025, F=2.243), CP1 electrode (p = 0.020, F=2.330), CP2 electrode (p = 0.011, F=2.539), CP6 electrode (p = 0.048, F=1.988), P7 electrode (p = 0.033, F=2.137), P3 electrode (p = 0.048, F=1.944), Pz electrode (p = 0.001, F=3.392), P4 electrode (p = 0.020, F=2.333), PO3 electrode (p = 0.014, F=2.466), PO4 electrode (p = 0.007, F=1.944, F=2.725), and O1 electrode (p = 0.030, F=2.178) in the central and parietal regions.

For absolute power values of the theta band, significant differences were found at the PO4 electrode (p = 0.044, F=2.032), P8 electrode (p = 0.038, F=2.090), and O1 electrode (p = 0.039, F=2.082).

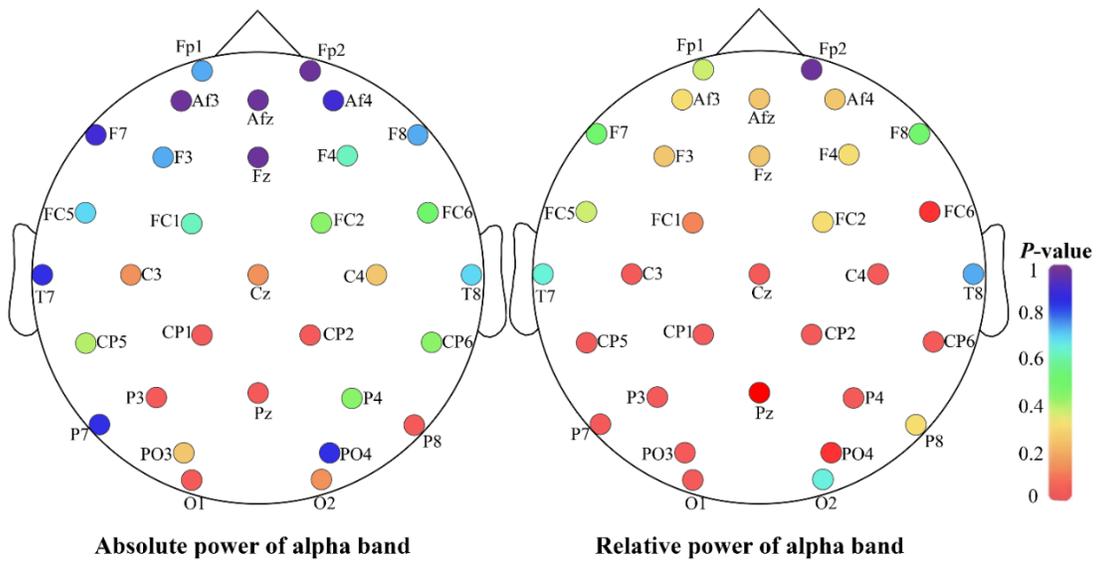

Fig.4. Significant brain dot distribution in left: absolute power of alpha band, right: relative power of alpha band.

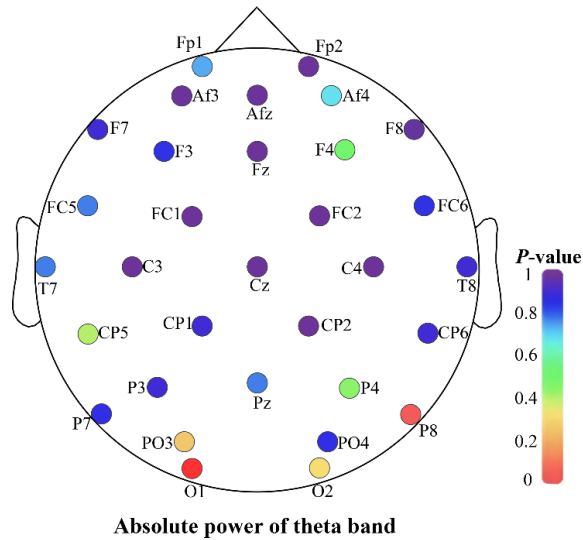

Fig.5. Significant brain dot distribution in absolute power of theta band.

### 3-3. Correlation of subjective measurements and EEG Data

Pearson correlation coefficients were computed to examine the relationship between significant channel power changes and subjective measurements, as illustrated in Fig. 6 and detailed in Table 5. The observations revealed that: (1) the absolute power of the alpha band exhibited a positive correlation with arousal and excitement ratings; (2) the relative power of the alpha band demonstrated a negative correlation with pleasantness, spaciousness, and interesting ratings.

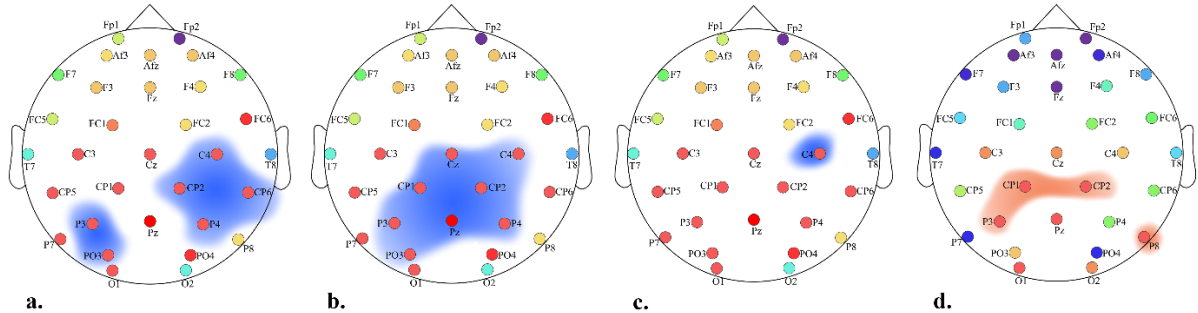

Fig. 6. The active channels correlated to the significant subjective measurements.
a. averaged alpha relative power topological graph across subjects under pleasantness ratings. Blue highlights indicate the scalp sites which show p<0.05.
b. averaged alpha relative power topological graph across subjects under spaciousness ratings. Blue highlights indicate the scalp sites which show p<0.05.
c. averaged alpha relative power topological graph across subjects under interest ratings. Blue highlights indicate the scalp sites which show p<0.05.
d. averaged alpha absolute power topological graph across subjects under excitement state. Red highlights indicate the scalp sites which show p<0.05.

**Table 5.** EEG channels with Pearson p-value <0.05 across different variables and Alpha band powers.

| AA×excitement | | | RA×pleasantness | | | RA×interest | | | RA×spaciousness | | |
|---|---|---|---|---|---|---|---|---|---|---|---|
| Ch. | PCC (r.) | p-value | Ch. | PCC (r.) | p-value | Ch. | PCC (r.) | p-value | Ch. | PCC (r.) | p-value |
| - | | | C4 | -.209** | .001 | C4 | -.150* | .016 | C4 | -.152* | .014 |
| CP2 | .199** | .003 | CP2 | -.158* | .011 | - | | | CP2 | -.176** | .004 |
| P3 | .220** | .001 | P3 | -.155* | .012 | - | | | P3 | -.148* | .016 |
| CP1 | .159* | .017 | - | | | - | | | CP1 | -.127* | .044 |
| P8 | .186** | .006 | - | | | - | | | - | | |
| - | | | P4 | -.145* | .018 | - | | | P4 | -.143* | .020 |
| - | | | PO3 | -.154* | .012 | - | | | PO3 | -.163** | .008 |
| - | | | CP6 | -.133* | .032 | - | | | - | | |
| - | | | - | | | - | | | Cz | -.130* | .037 |
| - | | | - | | | - | | | Pz | -.124* | .046 |

**.** Correlation is significant at the 0.01 level (2-tailed). *. Correlation is significant at the 0.05 level (2-tailed).

In Fig. 6a, the statistically significant increase in activation related to pleasant conditions is highlighted in blue. This result indicates that illumination levels perceived as highly pleasant elicited a lower activity of relative alpha power across parietal and central scalp sites. Fig. 6b emphasizes a broad increase of alpha activity across centro-parietal areas for spacious perception. Additionally, there is an increase of alpha activity at electrode C4 related to interesting conditions, and an increase of alpha activity across the central and right parietal regions associated with arousing and exciting illuminated conditions, as visible in Fig. 6c and 6d.

Due to the constraints of the article, a detailed analysis was conducted on three specific channels among the significant channels shown in Fig. 6 and Table 4:
- CP2 electrode: The relative alpha power of this electrode exhibits a negative correlation with pleasantness and spaciousness.
- C4 electrode: The relative alpha power of this electrode shows a negative correlation with pleasantness, interest, and perception of spaciousness in the space.
- P3 electrode: The absolute alpha power of this electrode demonstrates a positive correlation with excitement and arousal ratings.

### 3-3-2. Cp2 electrode

Considering the negative relationship between the variables of pleasantness and spaciousness with the relative power of the alpha band in the CP2 channel, its relationship was investigated among the nine groups of illumination levels, and the difference was significant (P=0.11, F=2.54). Post hoc comparisons using Tukey's HSD test revealed a significant difference between the average relative power of the alpha band in illuminance levels 3 (M=1.15, SD=0.6539) and 6 (M=1.193, SD=0.38) with level 9 (M=1.765, SD=0.99). As depicted in Fig. 7, since the low levels of Alpha relative power in the mentioned channel are associated with high levels of pleasantness and spaciousness, illuminance levels 3 (300-500 lux) and 6 (700-900 lux), with the lowest average relative power of the alpha band, have increased pleasantness and spaciousness compared to illuminance level 9 (1300-1500 lux).

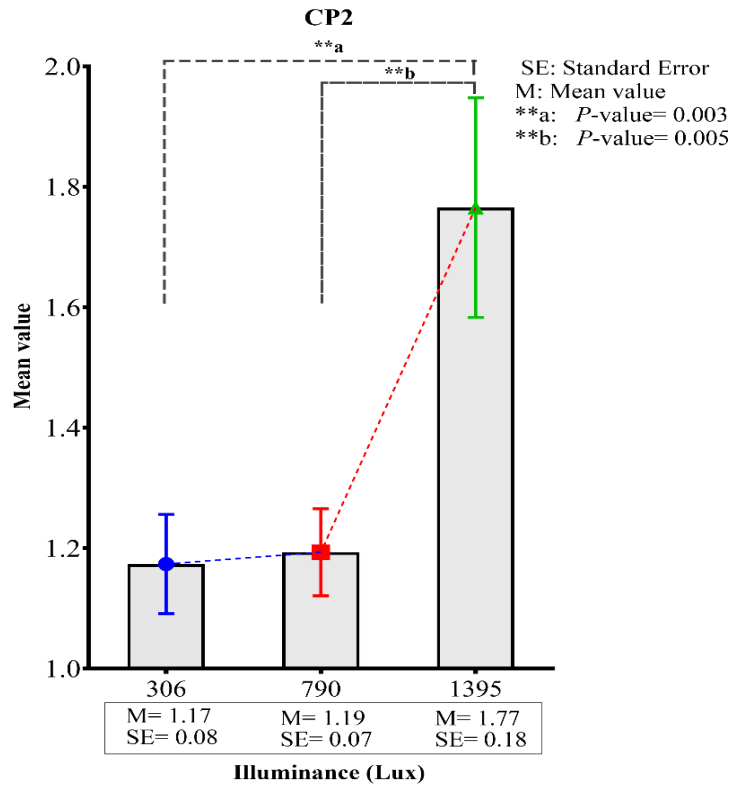

Fig. 7. Different CP2 relative Alpha means according to 3 illumination levels. The vertical line represents the standard error.

### 3-3-3. C4 electrode, interest, pleasantness and spaciousness:

In relation to the inverse correlation between the variables of pleasantness, interest, and spaciousness with the relative power of the alpha band in the C4 channel, an analysis of the relative power of the alpha band across nine illumination groups revealed a significant difference (P=0.01, F=2.566). Post hoc comparisons using Tukey's HSD test indicated a significant distinction in the average relative power of the alpha band in illuminance levels 2 (M=1.18, SD=0.406), level 3 (M=1.11, SD=0.443), and level 6 (M=1.20, SD=0.33) compared to level 9 (M=1.77, SD=1.03). Similar to channel CP2 and as illustrated in Fig. 8, illuminance levels 3 (300-500 lux) and 6 (900-700 lux), characterized by the lowest average relative power of the alpha band, not only enhanced pleasantness and spaciousness but also contributed to the perception of interest compared to illumination level 9 (1500-1300 lux). In this comparison, illumination level 2 was excluded based on the p-value and the aforementioned questionnaire results.

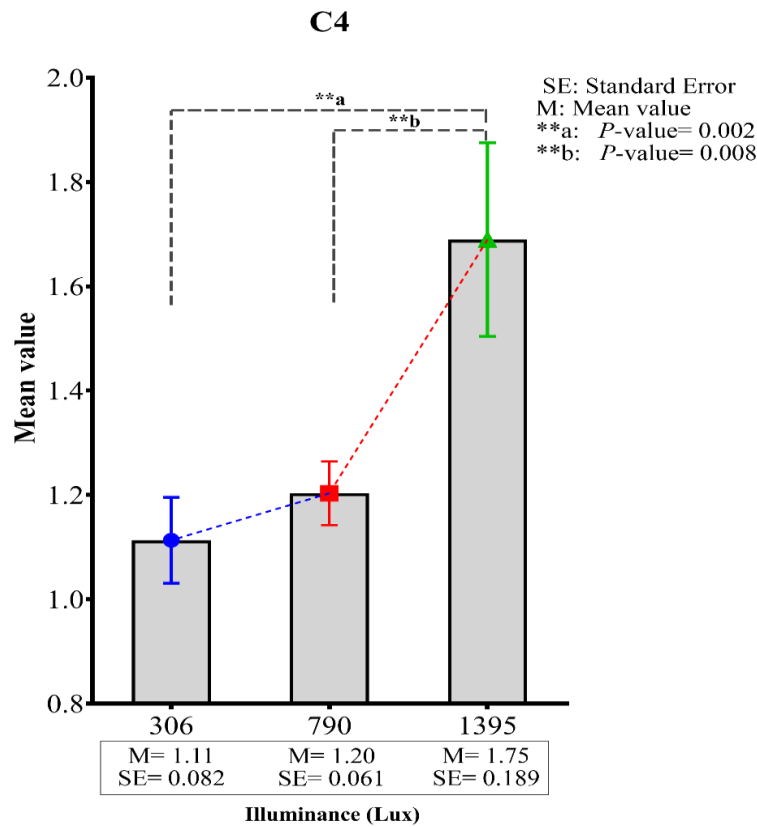

Fig. 8. Different C4 relative Alpha means according to 3 illumination levels. The vertical line represents the standard error.

### 3-3-1. P3 electrode

Considering the positive relationship between the excitement variable and the absolute power of the alpha band in the P3 channel, the relationship between them among nine levels of illumination has been investigated and the difference was significant (P=0.048, F =1.998). Post hoc comparisons using Tukey's HSD test showed that there is a significant difference between the average absolute power of the alpha band in the first illuminance level and level 9th. An independent samples t-test was performed to compare the absolute power of the alpha band in these two illuminance levels.

Statistically, there was a significant increase from 1(M=1.15, SD=0.6539) to 9 (M=2.06, SD=1.69) illumination levels. According to Fig. 9, since the low levels of the absolute power of the alpha band in the mentioned channel are related to the low level of excitement, the lighting level 1 (less than 100 lux) with the lowest average absolute power of the alpha band, cause a decrease in excitement compared to The illuminance level is 9 (1300-1500 lux).

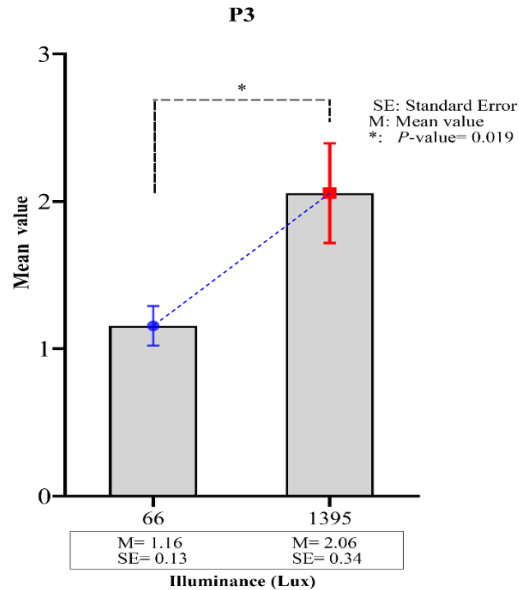

Fig. 9. Different P3 absolute Alpha means according to 2 illumination levels. The vertical line represents the standard error.

## 4. Discussion

The EEG analysis results reveal significant differences in band power at various task illuminance levels, particularly in channels situated in the parietal region, which is closely linked to human cognitive and perceptual activities (Gottlieb, 2007; Tong et al., 2023). This underscores the impact of changing daylight illumination levels on human cognitive and perceptual functions. In terms of EEG task absolute alpha band power, elevated values at P8, P3, CP2, and CP1 locations under high illumination conditions suggest that such conditions may induce heightened arousal, excitement, and reduced concentration (Shan et al., 2019). This interpretation aligns with previous findings that changes in alpha activation are associated with higher illuminance levels in the parietal lobe

On the contrary, the diminished value of relative alpha band power at P3, P4, PO3, C4, CP2, and CP6 locations under median and low illumination conditions suggests that these illumination levels might induce elevated pleasant emotions and heightened concentration. This interpretation aligns with previous findings that changes in alpha activation are associated with higher illuminance levels in the parietal lobe. This evaluation is consistent with previous reports suggesting that, changing alpha activation is correlated with higher illuminance levels in parietal lobe (Min et al., 2013; Armanto et al., 2009).

In summary, examining the correlation between the absolute and relative power of the alpha band in active channels with pleasantness and arousal variables suggests that illumination levels ranging from 300 to 900 lux can be considered pleasant. Additionally, an illuminance level exceeding 1300 lux appears to induce excitement and arousal, while levels below 100 lux may contribute to drowsiness. Considering the analysis of subjective ratings and the upward trend in pleasantness and arousal, the remaining illumination levels can be positioned in the 2D valence-arousal emotional model, as depicted in Fig. 10.

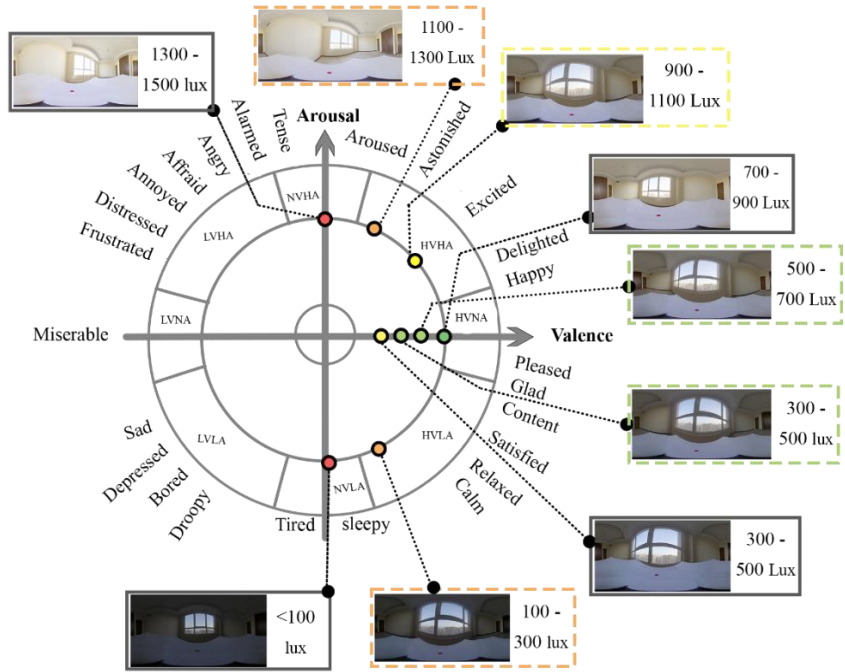

Fig. 10. Different task illumination levels and 2D valence-arousal emotion model by Russell (PAD model).

Regarding the impact of illuminance levels on architectural experience, participants reported variations in pleasantness, arousal, interest, and spaciousness corresponding to different illuminance levels. The combination of EEG and subjective evaluation analysis revealed distinctions in daylight illumination preferences across different states. Fig. 10 illustrates that the minimum illumination recommended for general workspaces is 300 lux, while 700 lux is suggested for more precise tasks. In general, a lighting range of 500 to 900 lux is considered suitable for office spaces, with an average of 700 lux for office rooms. This range not only fosters positive emotional experiences such as pleasantness and calmness in employees but also enhances their performance and concentration, while maintaining a balanced level of excitement.

As the illumination level increases, there is a corresponding increase in people's arousal and a decrease in concentration. This level of illumination can be suitable for spaces designed to be exciting, such as sports areas within organizations. However, spaces with lighting exceeding 1300 lux may not be ideal for prolonged employee stay due to high arousal and low pleasantness. On the other hand, illumination levels below 300 lux result in decreased pleasantness and arousal, and lighting below 100 lux may induce drowsiness.

The findings align with previous studies, such as Fu et al. (Fu et al., 2023), which suggest that 300 lux is suitable for rest spaces. Similar observations are reflected in artificial lighting studies, including Boyce and Cuttle's research indicating that higher lighting levels (600 lux) contribute to a more pleasant and comfortable environment (Boyce & Cuttle, 1990). Smulders et al. found that participants feel more alert and alive in 1000 lux conditions compared to 200 lux conditions, reducing sleepiness and increasing excitement (Smolders et al., 2012). Fang et al. demonstrated in their study that increasing illumination from 300 to 500 lux in the work environment decreases fatigue (Fang et al., 2022). Chao et al. identified lighting levels of 400 to 900 lux as bright and pleasant, 300 to 900 lux as comfortable, and 600 to 900 lux as an aroused environment (Chao et al., 2020). Moreover, the results are consistent with previous research recommending optimal lighting for accurate tasks to be above 500 lux (Chen et al., 2022; Tong et al., 2023; C.-C. Lin 2014; Wu et al. 2022; Fu et al. 2022).

### 4-1. Limitations:

While our study has offered valuable insights, certain limitations need acknowledgment. The display constraints of virtual reality glasses resulted in the omission of crucial daylight-related indicators, such as glare, which are significant in a real office setting. Although assessing this indicator with current virtual reality glasses is nearly impossible, it presents an avenue for future studies. The methodology applied in this study could be extended to examine existing lighting standards in other environments, including educational spaces. Additionally, gender and age differences were not considered in this study, which can be assessed in future research.

The subjective experiences in architectural design encompass various sensory modalities and go beyond the scope of EEG signals alone. For a more comprehensive understanding of these experiences, future investigations should emphasize the integration of EEG signals with data from diverse modalities, including eye tracking and environmental sensors. Such an integrated approach would provide a more holistic insight into subjective experiences in architectural design tasks. Furthermore, decoding EEG patterns for architectural design is a complex endeavor that necessitates collaboration between neuroscientists, architects, and designers. Future research can facilitate improved collaboration, enabling these professionals to gain deeper insights into how our brains respond to different environments.

### 5. Conclusion

In this study, we explored the electroencephalographic activation associated with the perception of a virtual office at different illuminance levels. Our hypothesis was that variations in illuminance could activate distinct cerebral circuits. Additionally, we aimed to determine the level of illuminance influencing architectural experiences such as pleasantness, calmness, and excitement, as well as how employees perceive their offices in terms of spaciousness, complexity, and interest, utilizing both subjective and objective measurements. The EEG and subjective ratings were recorded across nine office spaces illuminated from 66 to 1500 lux. The results revealed diverse effects of illuminance levels on various spatial experiences, which can be categorized as follows:

**Emotional Experience:**
- Illumination ranging from 300 to 900 lux is perceived as pleasant by employees, with an increase in pleasantness as illuminance level rises. The most pleasant condition is observed in the range of 700 to 900 lux.
- Illuminance levels exceeding 1300 lux led to increased arousal and excitement in individuals.
- Illumination below 300 lux reduces arousal and promotes calmness in employees, while levels below 100 lux may induce sleep.

**Neurophysiological Experience:**
- Illumination level significantly influences the power of the alpha and theta bands.

- Changes in daylight illumination levels activate the parietal and central regions of the brain associated with cognitive functions.
- Within the illumination range of 300 to 900 lux, there is a significant decrease in the relative power of the alpha band in the parietal and central brain regions.
- Illuminance levels exceeding 1300 lux result in a significant increase in the absolute power of the alpha band.
- Illumination levels below 300 lux led to a reduction in the absolute power of the alpha band.

## 6. Acknowledgements

All of the experiment sessions took place in the National Brain Mapping Laboratory (NBML).
**Data Availability Statement**: data is available at this link: https://drive.google.com/folderview?id=1rIJgBuefc0wNNur2wFQBeiThBYxKG8qM
**Funding**: The authors declare that they don't receive any fund for current research.
**Conflicts of interest/Competing interests:** The authors declare that they have no conflicts of interest.
**Ethics approval**: The experimental protocols were approved by Research Ethics Committees of Iran University of Medical Sciences (IR.IUMS.REC.1401.1033).
**Consent for publication:** N/A
**Authors contribution:** Pegah Payedar-Ardakani: Designed the analysis, Performed the analysis, Wrote the paper. Yousef Gorji-Mahlabani: Designed the analysis, proofreading. Abdolhamid Ghanbaran: Designed the analysis, proofreading. Reza Ebrahimpour: Designed the analysis, proofreading.